\documentclass[journal,10pt,twocolumn,twoside]{IEEEtran}

\usepackage{cite}
\usepackage{amsmath,amssymb,amsfonts}
\usepackage{algorithmic}
\usepackage{graphicx}
\usepackage{subfigure} 
\usepackage{textcomp}
\usepackage{booktabs}
\usepackage{tabularx}
\usepackage{caption}

\usepackage{amsthm}

\newtheorem{proposition}{Proposition}

\begin{document}

\title{Memory-Augmented Generative AI for Real-time Wireless Prediction in Dynamic Industrial Environments}

\author{Rahul Gulia, \IEEEmembership{Student Member, IEEE}, Amlan Ganguly, \IEEEmembership{Senior Member, IEEE}, Michael E. Kuhl, \IEEEmembership{Senior Member, IEEE}, Ehsan Rashedi, \IEEEmembership{Senior Member, IEEE}, and Clark Hochgraf, \IEEEmembership{Senior Member, IEEE}}

\markboth{Vol. XX, No. XX, 2026}
{Gulia \MakeLowercase{\textit{et al.}}: Memory-Augmented Generative AI for Wireless Prediction}

\maketitle

\begin{abstract}
Accurate and real-time prediction of wireless channel conditions, particularly the Signal-to-Interference-plus-Noise Ratio, is a foundational requirement for enabling Ultra-Reliable Low-Latency Communication in highly dynamic Industry 4.0 environments. Traditional physics-based or statistical models fail to cope with the spatio-temporal complexities introduced by mobile obstacles and transient interference inherent to smart warehouses. To address this, we introduce Evo-WISVA (Evolutionary Wireless Infrastructure for Smart Warehouse using VAE), a novel synergistic deep learning architecture. This framework ensures real-time foresight into wireless performance, supporting advanced industrial automation and proactive resource management. Evo-WISVA integrates a memory-augmented Variational Autoencoder featuring an Attention-driven Latent Memory Module for robust, context-aware spatial feature extraction, with a Convolutional Long Short-Term Memory network for precise temporal forecasting and sequential refinement. The entire pipeline is optimized end-to-end via a joint loss function, ensuring optimal feature alignment between the generative and predictive components. Rigorous experimental evaluation conducted on a high-fidelity ns-3-generated industrial warehouse dataset demonstrates that Evo-WISVA significantly surpasses state-of-the-art baselines, achieving up to a 47.6\% reduction in average reconstruction error. Crucially, the model exhibits exceptional generalization capacity to unseen environments with vastly increased dynamic complexity (up to ten simultaneously moving obstacles) while maintaining amortized computational efficiency essential for real-time deployment. Evo-WISVA establishes a foundational technology for proactive wireless resource management, enabling autonomous optimization and advancing the realization of resilient industrial communication networks.
\end{abstract}
\vspace{-3mm}

\vspace{-3mm}


\section{Introduction}
\IEEEPARstart{T}{he} evolution of wireless communication (5G, 6G) necessitates accurate and reliable prediction of wireless channel conditions \cite{Intro_1_SCNN_LSTM, Intro_7}. The Signal-to-Interference-plus-Noise Ratio (SINR) is a foundational metric crucial for applications demanding ultra-low latency, such as smart cities, industrial IoT, and connected autonomous systems. Forecasting SINR is essential for proactive network planning, dynamic resource allocation, and maintaining system reliability \cite{Intro_1_SCNN_LSTM, Intro_7}. In constrained settings like WSNs, advanced prediction also supports energy conservation \cite{Inro_9, Intro_10}.

Accurate SINR prediction is challenging, particularly in dynamic industrial environments \cite{Intro_2, Intro_3, Intro_4, Intro_5}. These settings combine deterministic factors (topology, materials) with highly stochastic elements (mobile obstacles, transient interference, multipath fading) \cite{Intro_2, Intro_3, Intro_4, Intro_5}. This complexity often yields noisy or incomplete data, making reliable forecasting difficult. Imperfect information leads to inefficient resource use and degraded quality of service. Mission-critical applications require trustworthy forecasting of link reliability, demanding more than simple point predictions \cite{Intro_2, Intro_3, Intro_4, Intro_5}.

Deep learning (DL) offers a transformative solution, leveraging its ability to learn complex, nonlinear spatio-temporal mappings \cite{Intro_6}. While Convolutional Neural Networks (CNNs) capture spatial features and Long Short-Term Memory (LSTM) networks capture temporal dependencies, existing architectures have limitations in dynamic RF environments \cite{Intro_6}. CNNs lack memory, and recurrent models often struggle with robustness and uncertainty disentanglement over long prediction horizons \cite{Intro_6}. Furthermore, real-time deployment requires efficient models capable of handling continuous data streams and incremental adaptation \cite{Intro_8, Intro_1_SCNN_LSTM, Intro_10}.

These challenges are most acute in Industry 4.0 environments, where mobile obstacles and automated logistics necessitate real-time foresight for Ultra-Reliable Low-Latency Communication (URLLC). To address this, we propose the Evolutionary WISVA (Evo-WISVA) model. This memory-augmented generative architecture is designed for spatio-temporal SINR forecasting, functioning as a neural surrogate capable of emulating the wireless environment in real time. By anticipating the evolution of the network state based on continuous measurements and environmental descriptors, Evo-WISVA enables proactive resource management and system-level optimization. This shifts wireless system design from a reactive to a predictive paradigm, advancing the realization of fully autonomous industrial wireless networks.

The potential impact of this work is substantial. By providing accurate SINR predictions, Evo-WISVA empowers new forms of wireless optimization, including dynamic beamforming, proactive handover, and intelligent AGV routing to avoid predicted dead zones. These capabilities enhance operational efficiency, reliability, and resilience within automated warehouses, reducing downtime and accelerating the realization of Industry 4.0. Importantly, the ability to emulate complex wireless effects quickly, rather than relying solely on computationally expensive simulation, represents a paradigm shift for industrial wireless connectivity.

This paper addresses the above challenges by introducing Evo-WISVA, a novel Memory-Augmented Real-time Generative AI model for wireless prediction. Our architecture provides accurate, robust, and timely spatio-temporal SINR forecasts, functioning as a neural surrogate of the physical wireless environment. Extensive experimental results demonstrate that Evo-WISVA significantly outperforms existing methods in predictive accuracy and efficiency within dynamic industrial environments. The following key novelties distinguish our approach:

\begin{enumerate}
    \item \textbf{Unified generative and predictive spatio-temporal modeling:} A novel integration of a memory-augmented Variational Autoencoder (VAE) for spatial feature learning and reconstruction, jointly trained with a Convolutional LSTM (ConvLSTM) for temporal sequence prediction.
    \item \textbf{Memory-augmented VAE for contextual spatial representation:} Introduction of a Latent Memory Module with attention in the VAE's latent space, enabling reconstructions that incorporate historical context for more coherent and robust SINR maps.
    \item \textbf{Hierarchical fusion and self-refining prediction:} A two-stage fusion mechanism where raw input maps are first fused in the VAE encoder, and the VAE's memory-augmented SINR reconstruction is then concatenated with the original inputs to enhance ConvLSTM predictions.
    \item \textbf{End-to-end trainability for feature alignment:} Unlike many hybrid models, the full pipeline (VAE + Latent Memory Module + ConvLSTM) is trained jointly with a combined loss function, ensuring optimal feature alignment between the generative and predictive components for superior SINR forecasting.
    \item \textbf{Domain-specific adaptation to dynamic RF environments:} The architecture is explicitly tailored for non-linear, time-varying wireless propagation in industrial settings with mobile obstacles, surpassing the limitations of static or statistical models.
\end{enumerate}

The remainder of this paper is structured as follows: Section~2 reviews related work on spatio-temporal deep learning and wireless channel modeling. Section~3 details the proposed Evo-WISVA architecture. Section~4 elaborates on the design novelties of our approach. Section~5 presents the experimental setup, results, and discussion. Section~6 concludes with insights and future research directions.


\vspace{-3mm}
\section{Related Work}
\vspace{-2mm}
Artificial Intelligence (AI) and Machine Learning (ML), particularly Deep Learning (DL), are transforming wireless communication by enabling systems to learn, reason, and optimize complex network functions \cite{Related_Work_1}. The accurate and real-time prediction of wireless channel conditions, specifically Signal-to-Interference-plus-Noise Ratio (SINR) maps, is crucial for next-generation network management and the development of advanced industrial communication networks. This section reviews channel prediction methodologies, highlighting their limitations and how our work, including the proposed Evo-WISVA model, addresses them.

\vspace{-3mm}
\subsection{Traditional Wireless Channel Modeling and Prediction}
Historically, wireless channel prediction relied on empirical, statistical, and deterministic models. Empirical models like the Log-Distance Path Loss Model offer simplicity but lack accuracy in complex, dynamic indoor environments \cite{Related_Work_2}. Our prior work, specifically the development of the UAV Low Altitude Air to Ground (U-LAAG) model, exemplified this, proposing a non-deterministic statistical model for indoor UAV/IoT wireless propagation in the 2.4–2.5 GHz ISM band to overcome the limitations of simpler empirical models in varied indoor settings \cite{Related_Work_3}.

Our subsequent research extensively investigated the challenges of 60 GHz millimeter-wave (mmWave) wireless connectivity within automated Industry 4.0 warehouses \cite{Related_Work_5, Related_Work_6}. These environments pose significant hurdles due to numerous non-line-of-sight (nLOS) paths caused by metallic shelves and storage boxes. The dynamic nature of warehouse configurations, combined with multipath reflections and shadow-fading, makes establishing stable, high-speed network coverage particularly difficult. Our initial Network Simulator-3 (NS-3) simulations demonstrated that 60 GHz network performance critically depends on LOS/nLOS existence, environmental reflectivity, and the density of autonomous material handling agents (AMHAs) \cite{Related_Work_5}. Notably, nLOS SINR was found comparable to LOS due to strong multipath components from metallic structures. Our expanded work further explored the impact of varying Access Point (AP) heights and different storage materials (wood, glass) on SINR, providing detailed SINR heatmaps to visualize connectivity transitions across warehouse aisles and confirming stable 60 GHz networks despite dynamic shelf configurations \cite{Related_Work_6}. These studies underscored that traditional channel models, and even advanced simulations, struggle with the real-time, site-specific, and dynamic complexities of industrial mmWave propagation, necessitating more adaptive prediction strategies \cite{Related_Work_7, Related_Work_8, Related_Work_9}.

Deterministic methods like ray-tracing, while accurate for static scenarios, are computationally prohibitive for real-time predictions in highly dynamic environments where obstacles are constantly in motion and material properties may be unknown \cite{Related_Work_10}. Collectively, traditional approaches lack the adaptive capabilities required for continuously evolving wireless conditions.

\vspace{-2mm}
\subsection{Machine Learning for Spatio-Temporal Wireless Channel Prediction}
\vspace{-1mm}
ML and DL offer powerful tools to model complex, non-linear relationships in wireless data, enabling adaptability beyond traditional methods \cite{Related_Work_11}. They are increasingly applied for Link Quality Prediction (LQP), resource allocation, and positioning in next-generation networks \cite{Related_Work_12,Related_Work_13,Related_Work_14}. However, existing ML models for channel prediction often face limitations regarding data quality, generalizability to diverse environments, scalability, and crucially, a lack of explicit confidence or "worst-case" performance guarantees, making them less trustworthy for critical industrial applications \cite{Related_Work_12,Related_Work_15,Related_Work_16,Related_Work_17,Related_Work_18}.

To address these challenges, particularly the need for efficient and automated wireless network design in dynamic industrial settings, our work introduced the Wireless Infrastructure for Smart warehouses using VAE (WISVA) model \cite{Related_Work_19}. WISVA leverages Variational Autoencoders (VAEs) for indoor radio propagation modeling within automated Industry 4.0 warehouses, focusing on 5G wireless bands. WISVA integrates physics-based tensors (e.g., permittivity, geometry, AP layout) via a three-branch encoder, enabling enhanced modeling of electromagnetic wave interactions. The model is trained on NS-3 simulated SINR heatmaps, learning to generate these heatmaps from latent tensors that encapsulate diverse warehouse configurations. This approach aims to simplify wireless network design by enabling the prediction of SINR heatmaps for both known and previously unseen warehouse layouts, including tasks like denoising and extrapolation, outperforming traditional autoencoders and SDU-Net models \cite{Related_Work_19}. This moves beyond resource-intensive direct measurements or traditional simulations.

Generative models, particularly Variational Autoencoders (VAEs), have shown promise in learning compact latent representations and capturing uncertainty in high-dimensional data \cite{Related_Work_23,Related_Work_24}. Our earlier WISVA model demonstrated that VAEs can generate SINR heatmaps directly from physics-informed tensors, reducing reliance on costly simulations or exhaustive measurements. However, standard VAEs process data largely frame by frame and fail to capture long-term temporal dependencies, limiting their utility for spatio-temporal forecasting.

To address these gaps, we propose Evo-WISVA, a memory-augmented generative architecture that functions as a data-driven neural surrogate of the wireless environment. Evo-WISVA integrates an attention-driven latent memory module within a VAE–ConvLSTM pipeline, enabling it to retain long-range temporal context while producing robust, denoised, and context-aware spatial features. Unlike models that merely mirror current conditions, Evo-WISVA anticipates the \emph{evolution} of SINR maps over time, providing real-time foresight into channel dynamics. This capability is particularly suited for Industry 4.0 scenarios, where wireless reliability must be continuously ensured despite moving obstacles, shifting layouts, and fluctuating interference. In doing so, Evo-WISVA closes the gap between generative spatial modeling and temporal forecasting, establishing a new class of lightweight, map-centric prediction tools for proactive network management.


\vspace{-2mm}
\section{Proposed Solution: Synergistic Deep Learning Architecture}
\begin{figure}[!t] 
\centering
\includegraphics[width=1.0\columnwidth]{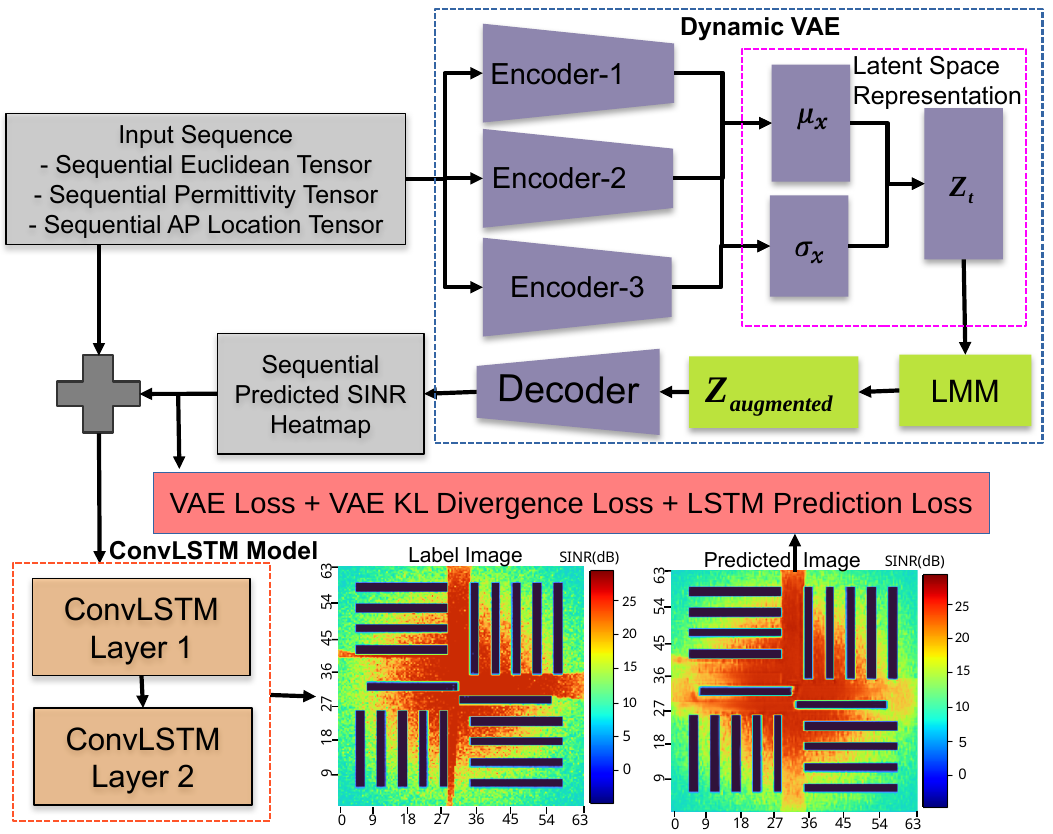}
\vspace{-2mm} 
\caption{Synergistic Deep Learning Architecture for Dynamic Wireless Prediction. }
\label{fig:architecture}
\vspace{-4mm} 
\end{figure}

\vspace{-1mm}
Our proposed solution, termed the Evolutionary WISVA (Wireless Infrastructure for Smart Warehouse using VAE) model (Evo-WISVA), represents an end-to-end trainable deep learning model meticulously engineered for accurate and efficient spatio-temporal Signal-to-Interference-plus-Noise Ratio (SINR) prediction within smart warehouses. This architecture is founded upon the synergistic integration of two primary interacting deep learning components: a novel WISVA model and a Convolutional Long Short-Term Memory (ConvLSTM) model. This design leverages the strengths of generative modeling for robust spatial understanding and recurrent neural networks for intricate temporal pattern learning. The holistic architecture is visually represented in Figure~\ref{fig:architecture}. This synergistic architecture allows Evo-WISVA to generate high-fidelity SINR heatmaps in real time, bridging physical warehouse layout changes and wireless performance forecasts.

\vspace{-2mm}
\subsection{Dynamic\_VAE Component: Memory-Augmented Spatial Understanding}
\vspace{-1mm}
The Dynamic\_VAE serves as the architecture's generative backbone, designed to process the input sequence $x_{seq}$ time step by time step. Its primary function is to distill a robust spatial understanding from diverse input modalities, dynamically enhanced by historical contextual information. For each time step $t$, the input $x_t \in \mathbb{R}^{\text{Batch} \times C \times H \times W}$ comprises $C=3$ distinct spatial maps, typically representing:
\begin{itemize}
    \item Euclidean Distance Map: Reflects the geometric proximity to Access Points (APs).
    \item Permittivity Map: Captures the material properties of the environment, influencing signal attenuation.
    \item AP Location Map: Indicates the positions of active APs.
\end{itemize}

These modalities are critical as they provide a comprehensive representation of the static and dynamic environmental factors that govern radio frequency (RF) propagation. The Dynamic\_VAE processes these inputs through the following four stages:

\paragraph{Encoder Module} The input $x_t$ is fed into the encoder $f_{enc}(\cdot)$, which employs parallel convolutional layers dedicated to specific input modalities. This parallel processing facilitates modality-specific feature extraction, capturing the unique physical influences of each map on RF propagation. The extracted features are concatenated and processed by shared layers to output the mean ($\mu_t$) and log-variance ($\log\sigma_t^2$) of the latent distribution for the current time step $t$:
\begin{align}
    \mu_t &= f_{enc, \mu}(x_t), \quad \log\sigma_t^2 = f_{enc, \log\sigma}(x_t) \label{eq:mu_t}
\end{align}

This encoding process acts as a nonlinear spectral filter. Under the ELBO objective $\mathcal{L}_{VAE} = \mathbb{E}_{q(z|x)}[\log p(x|z)] - \beta D_{KL}(q(z|x) || p(z))$, the latent bottleneck $z$ identifies a $d$-dimensional manifold $\mathcal{Z}$ spanning the dynamic subspace. By constraining the latent space to a standard Gaussian prior, the VAE filters dominant stationary eigen-directions ($\lambda_{static}$) that would otherwise induce gradient saturation in the temporal layers.

\paragraph{Reparameterization Trick} To enable backpropagation through the stochastic sampling, a latent vector $z_t$ is sampled using the reparameterization trick \cite{Related_Work_21}:
\begin{equation}
    z_t = \mu_t + \sigma_t \odot \epsilon, \quad \epsilon \sim \mathcal{N}(0, I) \label{eq:reparameterization_trick}
\end{equation}
where $\odot$ denotes the element-wise product. This ensures that gradients flow deterministically to optimize the encoder parameters.

\paragraph{Latent Memory Module (LMM)} The vector $z_t$ is passed to the LMM to integrate historical context and enhance temporal consistency. The LMM maintains a memory bank $\mathcal{M}$ of past latent states. An attention mechanism selectively weighs information from $\mathcal{M}$ using $z_t$ as the query ($Q$) and past states as keys ($K$) and values ($V$). The weights $a_{t,j}$ for each entry $m_j \in \mathcal{M}$ are computed as:
\begin{equation}
    a_{t,j} = \text{softmax}\left( \frac{(z_t)^T m_j}{\sqrt{d_k}} \right) \label{eq:attention_weights}
\end{equation}
The augmented latent vector $z_{augmented\_t}$ is then derived as a weighted sum: $z_{augmented\_t} = z_t + \sum_{m_j \in \mathcal{M}} a_{t,j} m_j$. This process informs the reconstruction with historical context, mitigating temporal "flickering" across the sequence.

\paragraph{Decoder Module} Finally, $z_{augmented\_t}$ is fed into the decoder $f_{dec}(\cdot)$, which utilizes deconvolutional layers to upsample the representation back to the spatial dimensions of the input. The output is the reconstructed SINR map for the current time step:
\begin{equation}
    \hat{y}_t = f_{dec}(z_{augmented\_t}) \label{eq:decoder_output}
\end{equation}

For the entire sequence, the VAE produces a sequence $\text{vae\_pred\_seq} \in \mathbb{R}^{\text{Batch} \times \text{Sequence\_Length} \times 1 \times H \times W}$, alongside means ($\mu_{seq}$) and log-variances ($log\_var_{seq}$) required for the joint loss calculation.

\vspace{-3mm}
\subsection{ConvLSTM Model Component: Temporal Refinement and Prediction}
\vspace{-2mm}
The ConvLSTM model is strategically employed for its capacity to model spatio-temporal dependencies while preserving spatial coherence. It takes a hierarchically fused input sequence, $lstm\_input\_seq$, derived from both the original raw inputs and the VAE's memory-augmented predictions. Specifically, $lstm\_input\_seq$ is a channel-wise concatenation of the original input sequence $x_{seq}$ and the Dynamic\_VAE's predicted SINR sequence $vae\_pred\_seq$:

\vspace{-3mm}
\begin{equation}
    lstm\_input\_seq_t = \text{concat}(x_t, \hat{y}_t) \label{eq:lstm_input_concatenation}
\end{equation}
\vspace{-3mm}

This results in a 4-channel input at each time step ($C' = 4$): [Euclidean, Permittivity, AP Location, VAE's Predicted SINR], for $lstm\_input\_seq_t \in \mathbb{R}^{\text{Batch} \times C' \times H \times W}$.

The ConvLSTM processes this enhanced input sequence time step by time step. Unlike traditional LSTMs that use matrix multiplications, ConvLSTM cells incorporate convolutional operations within their internal gates (input gate $i_t$, forget gate $f_t$, output gate $o_t$) and cell state ($C_t$), making them uniquely suited for grid-like data like spatial maps. The core operations within a ConvLSTM cell are given by:

\vspace{-3mm}
\begin{align}
    i_t &= \sigma(W_{xi} * X_t + W_{hi} * H_{t-1} + W_{ci} \odot C_{t-1} + b_i) \label{eq:convlstm_it}\\
    f_t &= \sigma(W_{xf} * X_t + W_{hf} * H_{t-1} + W_{cf} \odot C_{t-1} + b_f) \label{eq:convlstm_ft}\\
    C_t &= f_t \odot C_{t-1} + i_t \odot \tanh(W_{xc} * X_t + W_{hc} * H_{t-1} + b_c) \label{eq:convlstm_ct}\\
    o_t &= \sigma(W_{xo} * X_t + W_{ho} * H_{t-1} + W_{co} \odot C_t + b_o) \label{eq:convlstm_ot}\\
    H_t &= o_t \odot \tanh(C_t) \label{eq:convlstm_ht}
\end{align}
\vspace{-3mm}

where $X_t$ is the input $lstm\_input\_seq_t$, $H_t$ is the hidden state, $C_t$ is the cell state, $*$ denotes convolution, $\odot$ denotes Hadamard product, $\sigma$ is the sigmoid function, and $W$ and $b$ are learnable convolutional filters and biases, respectively. This architecture allows the ConvLSTM to preserve spatial coherence while learning intricate temporal dependencies across the sequence. It learns to refine the SINR prediction by observing how the original environmental features evolve alongside the VAE's memory-augmented initial estimate, effectively performing a "self-refinement" process. The output of the ConvLSTM is $outputs\_seq \in \mathbb{R}^{\text{Batch} \times \text{Sequence\_Len} \times 1 \times H \times W}$, representing the final predicted SINR maps.

\subsection{Theoretical Justification: Spectral Stability and Operator Synergy}

To analyze the stability and predictive reliability of the Evo-WISVA architecture, we examine the spectral properties of the transition operator $\mathcal{T}$. This hybrid design explicitly addresses the fundamental trade-off between temporal gradient stability and spatial inductive bias inherent in industrial RF environments.

\begin{proposition}[LMM as a Non-Markovian Gradient Shortcut]
Let $H_t$ be the hidden state of the ConvLSTM refinement module. In traditional recurrent architectures, the sensitivity of the loss $\mathcal{L}$ to a state $k$ steps in the past is governed by the temporal Jacobian $J_t = \frac{\partial H_t}{\partial H_{t-1}}$. By the chain rule and the sub-multiplicative property of the spectral norm $\|\cdot\|_2$:
\begin{equation}
\left\| \frac{\partial \mathcal{L}}{\partial H_{t-k}} \right\|_2 \leq \left\| \frac{\partial \mathcal{L}}{\partial H_t} \right\|_2 \prod_{i=0}^{k-1} \sigma_{max}(J_{t-i})
\end{equation}
In high-mobility industrial tracking, $\sigma_{max}(J) < 1$ is required for state stability, which leads to vanishing gradients $\lim_{k \to \infty} \|\frac{\partial \mathcal{L}}{\partial H_{t-k}}\| = 0$ \cite{hochreiter1997long}. The Evo-WISVA \textbf{Latent Memory Module (LMM)} introduces an additive attention operator $\mathcal{A}$ such that the augmented latent state is $z_{augmented\_t} = z_t + \sum a_{t,j} m_j$. The gradient path for the memory component is:
\begin{equation}
\frac{\partial z_{augmented\_t}}{\partial m_j} = a_{t,j} I
\end{equation}
Because this term is independent of the sequence depth $k$, it ensures the effective spectral radius $\rho(J_{eff}) \approx 1$. This mathematical shortcut enables long-term temporal credit assignment for periodic industrial patterns that are unreachable by standard recurrent units.
\end{proposition}

\begin{proposition}[ConvLSTM as a Local Differential Operator]
While the LMM is permutation-invariant and lacks inherent spatial topology, the ConvLSTM module defines the transition as a spatial convolution: $H_t = \sigma ( \mathcal{K} * H_{t-1} + \mathcal{W} * \text{lstm\_input\_seq}_t + b )$. This operator satisfies the spatial continuity requirement derived from the transport equation $\nabla \cdot \vec{J} + \frac{\partial \rho}{\partial t} = 0$ \cite{Related_Work_2}, ensuring that the SINR value at coordinate $(x,y)$ is functionally dependent on its immediate neighborhood $\mathcal{N}(x,y)$. This enforces a physical smoothness constraint on the latent trajectory, modeling the "flow" of RF shadows as a continuous differential process.
\end{proposition}

\begin{proof}[Synergy and URLLC Convergence]
The Evo-WISVA architecture represents a hybrid Integral-Differential operator. The LMM serves as a \textbf{Global Integral Operator}, $h_{global} = \int \alpha(\tau) z(\tau) d\tau$, capturing long-range shadowing context. The ConvLSTM serves as a \textbf{Local Differential Operator}, $\frac{\partial H}{\partial t} \approx f(H, \nabla H)$, ensuring spatial topology and local coherence. Together, they satisfy the composite update:
\begin{equation}
H_{final} = \underbrace{\text{ConvLSTM}(H_{t-1}, \hat{y}_t)}_{\text{Spatial Locality Bias}} + \underbrace{\text{LMM}(\mathcal{M})}_{\text{Temporal Gradient Stability}}
\end{equation}
This dual-path optimization ensures the error $\epsilon$ remains bounded even for $T \gg 1$, satisfying the strict predictive reliability required for URLLC in dynamic warehouse environments \cite{Levie2021_RadioUNet}.
\end{proof}
\vspace{-3mm}

\subsection{Joint Loss Calculation and Optimization}
The entire architecture is optimized end-to-end using a single joint loss function. This unified strategy ensures synergistic learning, compelling both the generative capabilities of the Dynamic\_VAE and the predictive accuracy of the ConvLSTM to evolve in concert. The total objective $L_{Total}$ comprises three distinct terms:

\paragraph{VAE Reconstruction Loss ($L_{VAE\_recon}$)} This term quantifies the accuracy of the VAE's reconstructed SINR maps against the ground truth. It is calculated as the Mean Squared Error (MSE) between the predicted sequence ($vae\_pred\_seq$) and the true sequence ($y_{seq}$):
\begin{equation}
    L_{\text{VAE\_recon}} = \frac{\sum_{b, t, h, w} (\text{vae\_pred\_seq}_{b,t,h,w} - y_{\text{seq},b,t,h,w})^2}{N \cdot T \cdot H \cdot W}
    \label{eq:vae_recon_loss}
\end{equation}
where $N$, $T$, $H$, and $W$ represent batch size, sequence length, height, and width, respectively.

\paragraph{VAE KL Divergence Loss ($L_{KLD}$)} This regularization term ensures that the latent distributions $\mathcal{N}(\mu_t, \sigma_t^2)$ are pushed towards a standard normal prior $\mathcal{N}(0, I)$. This promotes a smooth, continuous, and disentangled latent space. For a multivariate Gaussian distribution, the KLD is formulated as:
\begin{equation}
    L_{KLD} = \frac{1}{2} \sum_{t=1}^{T} \sum_{i=1}^{D} (\sigma_{t,i}^2 + \mu_{t,i}^2 - 1 - \log(\sigma_{t,i}^2)) \label{eq:kld_loss_full}
\end{equation}
where $D$ is the latent dimension. This term penalizes deviations from the prior, preventing posterior collapse.

\paragraph{LSTM Prediction Loss ($L_{LSTM\_pred}$)} This term measures the final spatio-temporal predictive accuracy of the ConvLSTM. It is the MSE between the model outputs ($outputs\_seq$) and the true SINR maps ($y_{seq}$):
\begin{equation}
    L_{\text{LSTM\_pred}} = \frac{\sum_{b, t, h, w} (\text{outputs\_seq}_{b,t,h,w} - y_{\text{seq},b,t,h,w})^2}{N \cdot T \cdot H \cdot W}
    \label{eq:lstm_pred_loss_full}
\end{equation}

The total loss $L_{Total}$ is a weighted sum of these components, where $\alpha$, $\beta$, and $\gamma$ are hyperparameters determined via empirical tuning to balance generative stability with predictive precision:
\begin{equation}
    L_{Total} = \alpha L_{VAE\_recon} + \beta L_{KLD} + \gamma L_{LSTM\_pred} \label{eq:total_loss}
\end{equation}


\vspace{-3mm}
\section{Experimental Evaluation}
\vspace{-1mm}
This section presents a comprehensive evaluation of the Evo-WISVA architecture, detailing the experimental methodology, implementation specifics, and a thorough comparative analysis against several state-of-the-art baselines. We also assess the model's performance on limited time-series datasets and analyze its real-time prediction capabilities.

\vspace{-3mm}
\subsection{Experiment Methodology}
\vspace{-2mm}
Our experiments are designed to rigorously test the Evo-WISVA model's predictive accuracy and computational efficiency in a dynamic wireless environment. We utilize a synthetic dataset generated from high-fidelity Network Simulator-3 (ns-3) simulations of a smart warehouse, which includes dynamic obstacles such as moving shelves. The dataset is partitioned into a training set and a testing set using a chronological split with a 75\%-25\% ratio.

\textbf{Dataset Credibility and Validation:} The use of ns-3 for 60~GHz industrial modeling is well-supported by recent comparative studies. Specifically, the authors \cite{ns3_defense_1} cross-validated the ns-3 802.11ad/ay modules against a physical 60~GHz testbed, reporting an average error of only 2.43\% and a standard deviation of 1.33~dBm. Furthermore, our environment utilizes the Quasi-Deterministic (Q-D) channel model, which authors \cite{ns3_defense_2} demonstrated to be mathematically equivalent to deterministic ray-tracing in terms of 3D obstacle specification and LoS/NLoS accuracy. This ensures that our training data captures site-specific multipath and shadowing effects with high physical fidelity.

\begin{figure*}[!t]
    \centering
    \includegraphics[width=\textwidth]{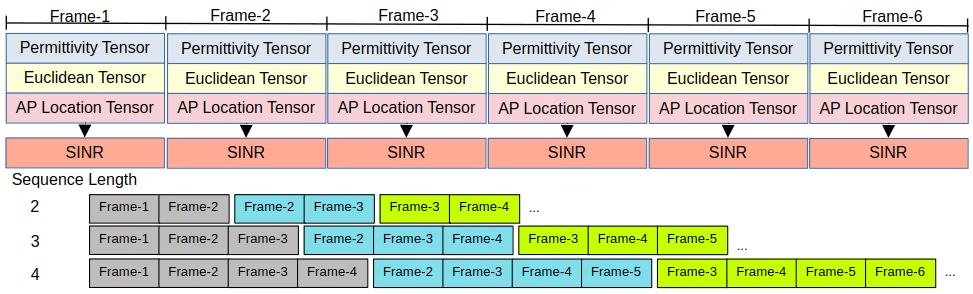}
    \caption{An illustration of the sliding window method used to create time-series sequences from the original dataset.}
    \label{fig:sequence_creation}
\end{figure*}

Following the chronological split, we transformed the datasets into time-series sequences using a sliding window approach. For a given `sequence\_length`, each input sample for the model consisted of a series of multi-channel images (Euclidean distance, Permeability, and AP location maps) corresponding to that time window. The corresponding label for each sequence was the series of SINR heatmaps for the same time steps. The window was slid by one time step at a time, generating overlapping sequences for training and evaluation, which allowed the model to learn the temporal dynamics between consecutive frames (Figure \ref{fig:sequence_creation}).

Performance is evaluated using standard regression metrics, including Mean Absolute Error (MAE) and Root Mean Square Error (RMSE).

\vspace{-3mm}
\subsection{Implementation Details}
\vspace{-1mm}
The Evo-WISVA architecture, including the Dynamic\_VAE with its Latent Memory Module and the ConvLSTM, is implemented using the \textbf{PyTorch} deep learning framework. All experiments are conducted on a server with an NVIDIA GPU (GeForce RTX 4090) to leverage parallel processing for efficient training and inference. The models are trained using the Adam optimizer with a learning rate of $1e-4$. We use the custom joint loss function described in Section 3 to optimize the VAE's generative capabilities and the ConvLSTM's predictive accuracy simultaneously.

\vspace{-3mm}
\subsection{Ablation Studies: Quantifying the Contribution of the LMM and the Hybrid Architecture}
\label{sec:Chapter-5_6_2}

\vspace{-3mm}
\begin{figure}[!h]
    \centering
    \includegraphics[width=1.0\columnwidth]{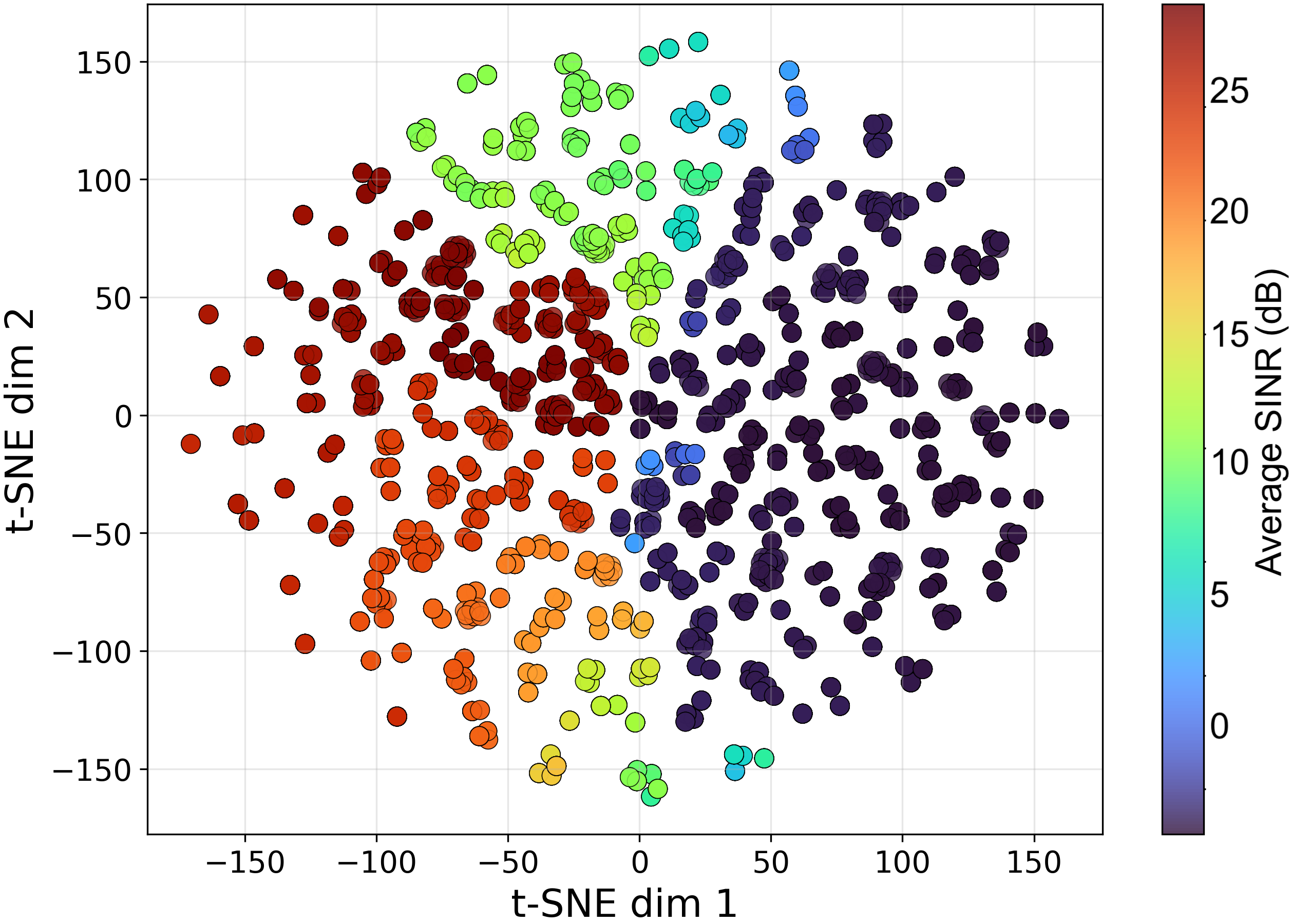}
    \caption{t-SNE of Evo-WISVA Latent Space}
    \label{fig:evo-wisva-tsne}
\end{figure}

\vspace{-3mm}
\begin{figure}[!h]
    \centering
    \includegraphics[width=0.9\linewidth]{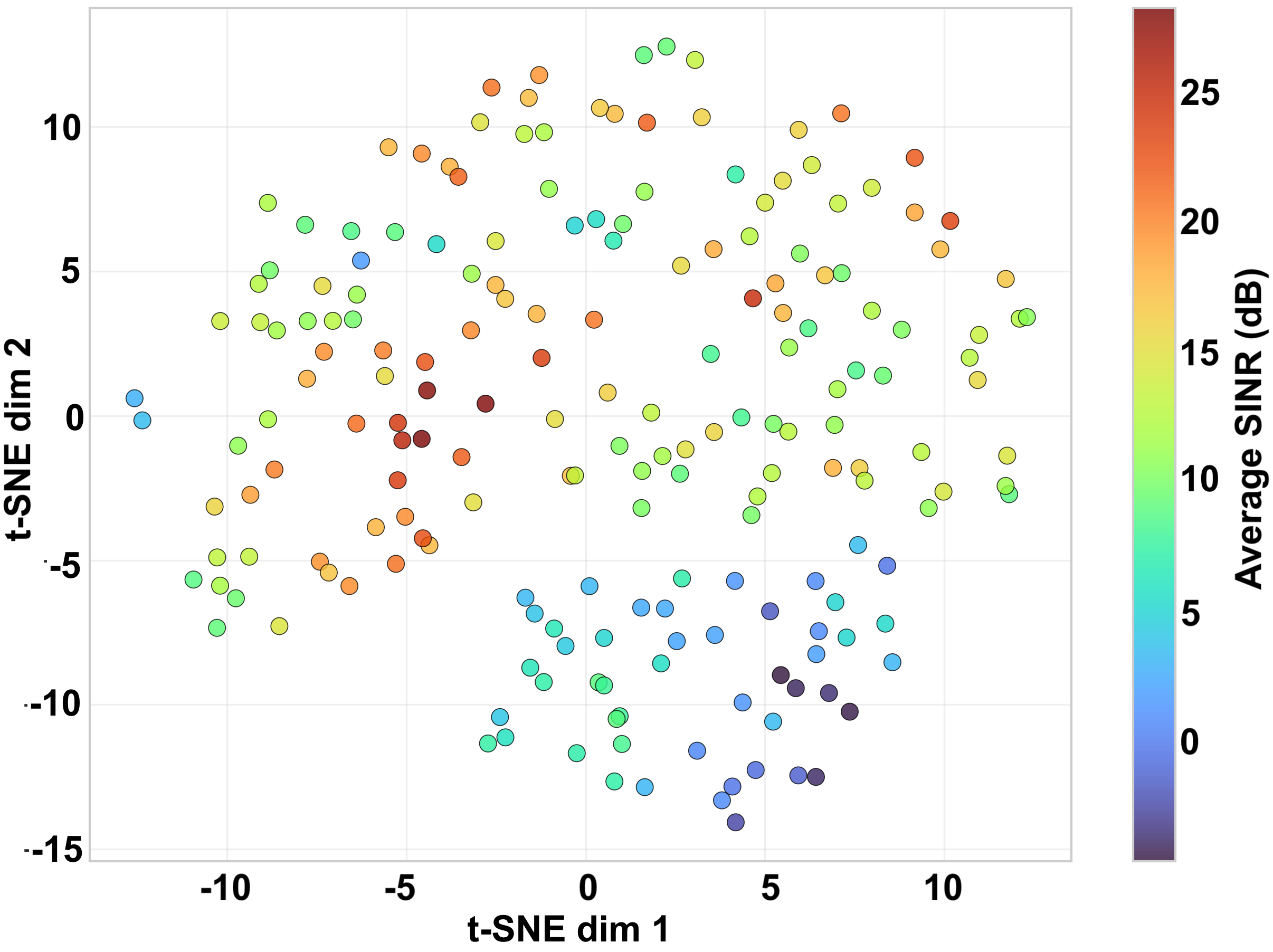}
    \caption{t-SNE of WISVA Latent Space.}
    \label{fig:WISVA_tSNE}
\end{figure}
\vspace{-3mm}

The comparison between the t-SNE visualizations of the Evo-WISVA latent space (Fig.~\ref{fig:evo-wisva-tsne}) and the baseline WISVA manifold (Fig.~\ref{fig:WISVA_tSNE}) provides crucial insight into the structural benefits of the \textbf{Attention-driven Latent Memory Module (LMM)} and ConvLSTM components. \textbf{Latent Space Scale and Separation:} A primary distinction is observed in the manifold scale; Evo-WISVA spans approximately $\pm 150$ on both axes, whereas WISVA spans only $\pm 15$. As a standard VAE, WISVA's latent vectors $\mathbf{z}$ are heavily constrained by the KL divergence loss, forcing adherence to a compact standard Gaussian prior $\mathcal{N}(0, \mathbf{I})$ and resulting in a denser, smaller-magnitude manifold. Conversely, the complex, memory-augmented Evo-WISVA architecture generates latent means $\boldsymbol{\mu}$ that are widely separated and less constrained by the strict Gaussian prior. The LMM and attention mechanism enforce significant distances between distinct contextual states, confirming the model's success in robustly encoding highly distinct environmental modes. 

\textbf{SINR Organization and Disentanglement:} Regarding feature organization, Evo-WISVA exhibits tight, highly homogeneous clusters that align monotonically along the second t-SNE dimension, with high SINR ($\approx 25$ dB) and low SINR ($\approx 0$ dB) concentrated at opposite poles. This precise differentiation contrasts sharply with WISVA, which presents a diffuse, scattered distribution with substantial cluster overlap and poor structural organization. The distinct clustering and clear ordering of the SINR metric in Evo-WISVA signify that the LMM and hybrid design successfully disentangle the SINR feature from other encoded environmental factors. This highly structured encoding is critical for the subsequent ConvLSTM to perform the accurate, context-aware temporal forecasting required for URLLC requirements. 

\textbf{Conclusion on Latent Quality:} These findings empirically validate Lemma 1, indicating that Evo-WISVA successfully expands the spectral representation of the dynamic SINR manifold. By filtering dominant static variance through the VAE and short-circuiting the temporal gradient via the LMM, the model achieves the disentanglement visible in Fig.~\ref{fig:evo-wisva-tsne}, where SINR levels are linearly separable along a single axis. This separation proves that the model has avoided the common pitfall of posterior collapse, preserving high-frequency spatial features necessary for industrial forecasting.

\begin{table}[htbp]
\centering
\caption{Sensitivity Analysis of Joint Loss Hyperparameters on Prediction Accuracy.}
\label{tab:sensitivity_analysis}
\begin{tabular}{@{}cccccc@{}}
\toprule
\textbf{Case} & \boldmath$\alpha$ \textbf{(VAE)} & \boldmath$\beta$ \textbf{(KLD)} & \boldmath$\gamma$ \textbf{(LSTM)} & \textbf{MAE (dB)} \\ \midrule
1 & 1.0 & 0.001 & 1.0 & \textbf{1.7451} \\
2 & 1.0 & 0.01  & 1.0 & 2.9091 \\
3 & 1.0 & 0.1   & 1.0 & 2.0498 \\
4 & 0.5 & 0.01  & 1.0 & 2.1149 \\
5 & 1.0 & 0.01  & 0.5 & 2.6189 \\
6 & 1.5 & 0.01  & 1.0 & 1.9485 \\
7 & 1.0 & 1.0   & 1.0 & 2.0590 \\ \bottomrule
\end{tabular}
\vspace{-2mm}
\end{table}

The sensitivity analysis, summarized in Table \ref{tab:sensitivity_analysis}, reveals that the Evo-WISVA framework is highly sensitive to the variational bottleneck weight $\beta$. Specifically, Case 1 ($\beta=0.001$) achieved the lowest error (MAE), suggesting that excessive KL-regularization in dynamic industrial environments can lead to ``posterior collapse,'' where subtle RF shadowing details are smoothed over. Furthermore, the performance gain from Case 4 to Case 6 demonstrates that the spatial reconstruction weight $\alpha$ is a primary determinant of accuracy, validating the VAE's role as a robust spatial prior. The ConvLSTM ($\gamma$) contributes to temporal refinement, though its efficacy is contingent upon the stability of the latent features provided by the VAE.

\vspace{-3mm}
\subsection{Evo-WISVA: Comparison with State-of-the-Art Models}
\vspace{-2mm}
To rigorously evaluate the efficacy of the proposed Evo-WISVA architecture, its performance was benchmarked against several state-of-the-art deep learning models for spatiotemporal wireless channel prediction. The selected baselines include a standalone ConvLSTM model \cite{SOTA_convLSTM}, a Wireless Infrastructure for Smart Warehouse using Variational Autoencoder (WISVA) model, an LSTM-based Link Quality Indicator (LSTM-LQI) model \cite{SOTA_LSTM_LQI}, and a GRU-LSTM model \cite{SOTA_LSTM_GRU}. Furthermore, to provide a broader context of architectural evolution, Table~\ref{tab:model_comparison} provides a comparative framework of these and other stochastic recurrent architectures, highlighting differences in memory design, spatial resolution, and physics retention. This comprehensive comparison is designed to highlight the advantages conferred by Evo-WISVA’s synergistic and memory-augmented design.

\begin{figure}[!h]
    \centering
    \includegraphics[width=1.0\columnwidth]{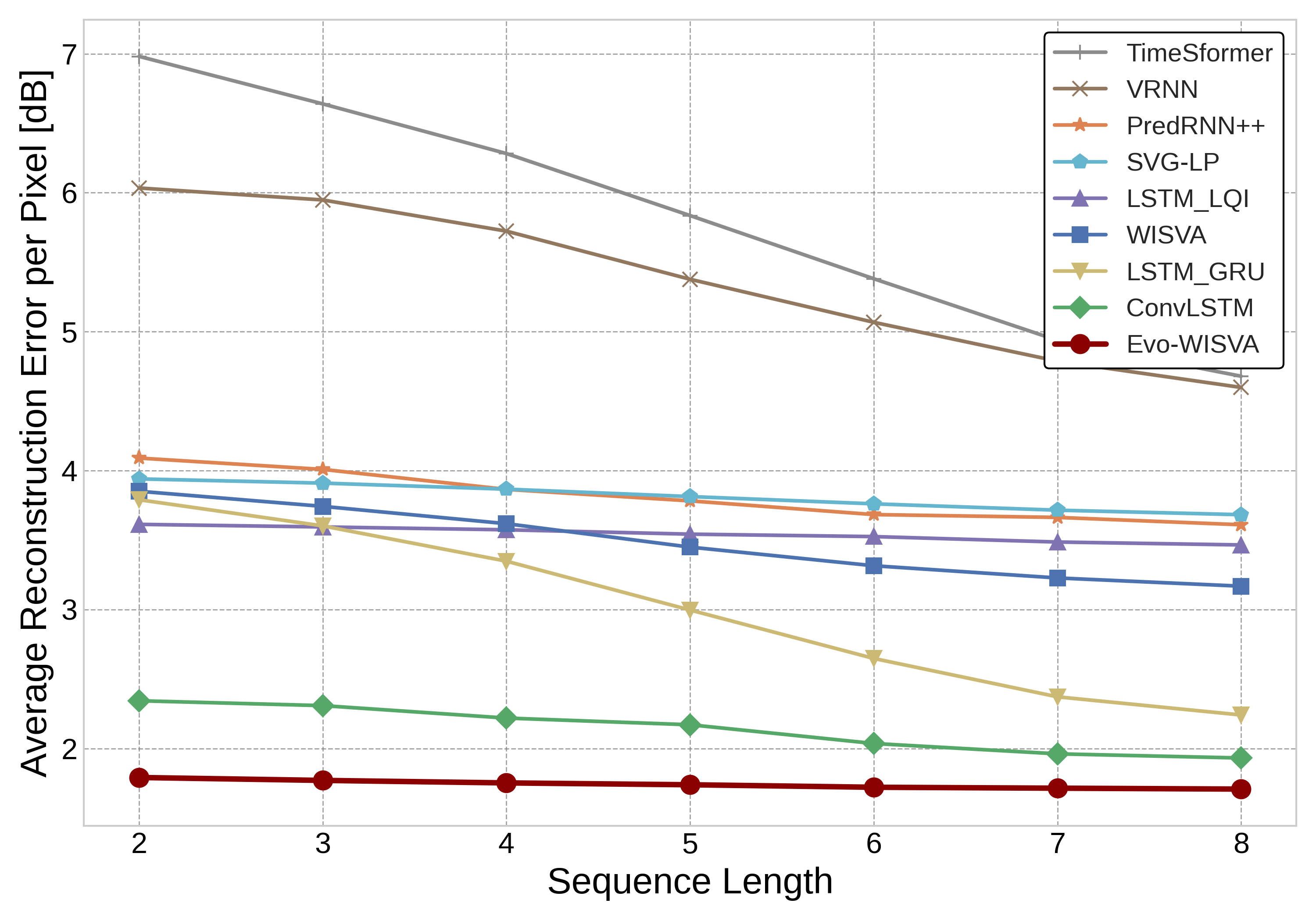}
    \caption{Average reconstruction error comparison of Evo-WISVA against various baseline models across different sequence lengths.}
    \label{fig:comparison}
\end{figure}

Figure~\ref{fig:comparison} illustrates the average reconstruction error per pixel (in dB) for each model as a function of the input sequence length. The data unequivocally demonstrate that the Evo-WISVA model consistently achieves a significantly lower prediction error across all tested sequence lengths. 

As synthesized in Table~\ref{tab:model_comparison}, the superior performance of Evo-WISVA is primarily attributed to its innovative hybrid architecture. While traditional models like SVG-LP \cite{SOTA_SVG-LP} or VRNN \cite{SOTA_VRNN} often suffer from hidden state drift or limited spatial resolution, the memory-augmented VAE component of Evo-WISVA provides a robust and context-aware initial estimate of the SINR map. 

The subsequent ConvLSTM module leverages this enriched spatiotemporal information to perform a precise refinement. This two-stage process—summarized by the ``Triple-Stream'' input logic in Table~\ref{tab:model_comparison}—effectively mitigates the inherent noise and stochasticity of the wireless channel, which simpler, single-stage models are less equipped to handle.

\begin{table*}[t]
    \centering
    \footnotesize 
    \setlength{\tabcolsep}{3pt} 
    \caption{Comparative Framework of Stochastic Recurrent Architectures for SINR Sequence Prediction}
    \label{tab:model_comparison}
    \begin{tabularx}{\textwidth}{@{} l *{5}{>{\raggedright\arraybackslash}X} @{} }
        \toprule
        \textbf{Model} & \textbf{Memory Architecture} & \textbf{Spatial Resolution} & \textbf{Input Streams} & \textbf{Temporal Dynamics} & \textbf{Physics Retention} \\
        \midrule
        SVG-LP \cite{SOTA_SVG-LP} & Hidden State only & Latent-only & Single Encoder & Learned Prior & Low (Drift) \\
        VRNN \cite{SOTA_VRNN} & Recurrent ($h_t$) & Latent + RNN & Coupled Encoder & Conditional Prior & Moderate \\
        LSTM-LQI \cite{SOTA_LSTM_LQI} & Dual LSTM (Det.+Stoch.) & Wavelet-Denoised SNR & Det.+Var. Streams & CI Boundary & Low (Point) \\
        Seq2Seq \cite{SOTA_LSTM_GRU} & Enc-Dec (LSTM/GRU) & Sequence RSSI & Single Signal & Recursive Unguided & Moderate \\
        ConvLSTM \cite{SOTA_convLSTM} & Recurrent ($C_t, H_t$) & Conv. Grid & Single (Radar/SINR) & Causal Conv. & Moderate \\
        WISVA \cite{Related_Work_17} & VAE Latent Space & Grid (152x152) & Multi-Physics & Static Recon. & High \\
        TimeSformer \cite{SOTA_TimeSformer} & Global Attention & Patch-based (ViT) & Frame Patches & Divided Space-Time & High (Global) \\
        PredRNN++ \cite{SOTA_PredRNN++} & Dual (C+M) + GHU & Multi-scale Conv & Cascaded Causal & Dual-Path Recur. & High (Long) \\
        \midrule
        \textbf{Evo-WISVA} & \textbf{Evol. Recur.+Attn.} & \textbf{Multi-Res ConvLSTM} & \textbf{Triple-Stream} & \textbf{Stoch. Attn. Prior} & \textbf{Ultra-High} \\
        \bottomrule
    \end{tabularx}
\end{table*}


\subsection{Visual Analysis of Spatio-Temporal Prediction Quality}
\begin{figure}[htbp!]
\centering 
\subfigure[Ground Truth SINR Map]{\label{fig:prediction_visualization:a}\includegraphics[height=35mm]{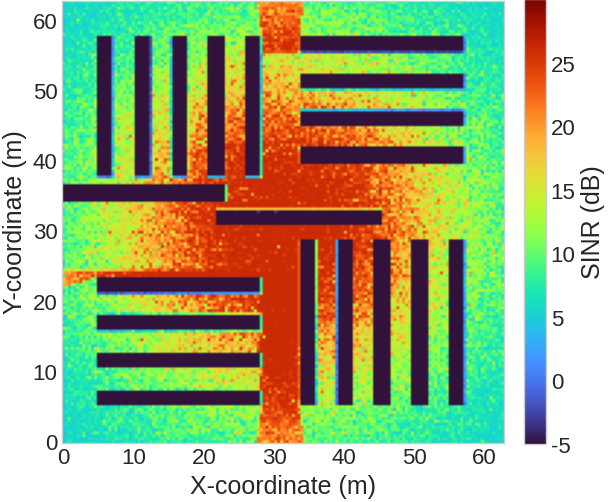}}
\hspace{1mm}
\subfigure[Evo-WISVA Predicted SINR Map]{\label{fig:prediction_visualization:b}\includegraphics[height=35mm]{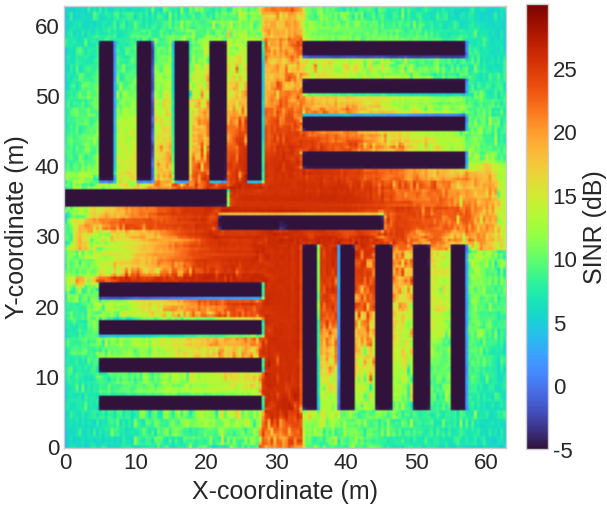}}
\hspace{1mm}
\subfigure[Reconstruction Error Map]{\label{fig:prediction_visualization:c}\includegraphics[height=37mm]{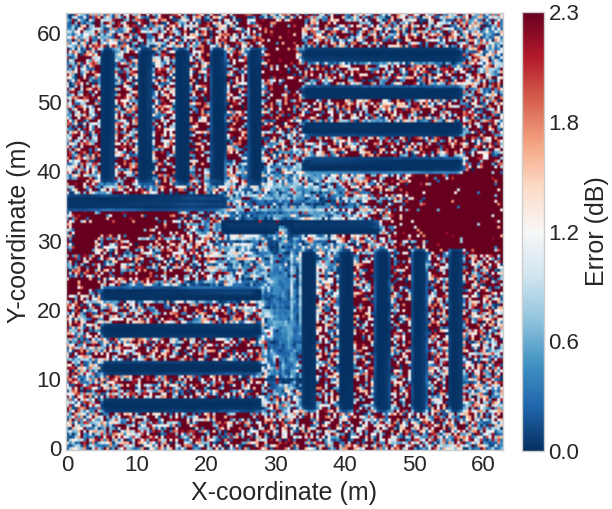}}
\caption{Visual comparison of Evo-WISVA's prediction at an unseen time step. Subfigure (a) is the input ground truth, (b) is the corresponding prediction, and (c) shows the pixel-wise reconstruction error, demonstrating high predictive fidelity across the dynamic environment.} 
\label{fig:prediction_visualization}
\end{figure}

\vspace{-1mm}
To qualitatively assess the Evo-WISVA architecture's ability to accurately capture the spatial distribution and temporal dynamics of the wireless channel, we present a visual comparison of the predicted SINR heatmap against the ground truth. Figure \ref{fig:prediction_visualization} showcases a randomly selected unseen frame from the test dataset, highlighting the model's predictive fidelity.

\vspace{-1mm}

A closer look at the visualizations (Figure \ref{fig:prediction_visualization}) directly supports the quantitative metrics. The predicted SINR field demonstrates strong spatial fidelity by accurately tracking the ground truth, effectively anticipating complex environmental effects like shadowing and multipath induced by dynamic obstacles. This accuracy is quantified by the reconstruction error map, which is overwhelmingly characterized by low errors (blue regions, near 0 dB), thus confirming the low Mean Absolute Error (MAE) established previously. Errors that are marginally higher (red regions) are localized to sharp boundaries and areas undergoing rapid line-of-sight (LOS) to non-line-of-sight (NLOS) transitions, which represent the most challenging predictive regions. In totality, this qualitative assessment reinforces the robustness and precision conferred by the novel memory-augmented, hybrid Evo-WISVA architecture.

\vspace{-3mm}
\subsection{Performance on Limited Time-Series Data}
\vspace{-2mm}
To rigorously assess the model's robustness and generalization capability, we conducted a performance comparison between its behavior on the entire validation dataset and on a limited subset of 100 unique sequential time-series data points. This experiment simulates a real-world deployment scenario where the model must perform reliably with a reduced volume of continuous data.

\begin{figure}[!t] 
    \centering
    \begin{minipage}{1.0\columnwidth}
        \centering
        \includegraphics[width=0.95\linewidth]{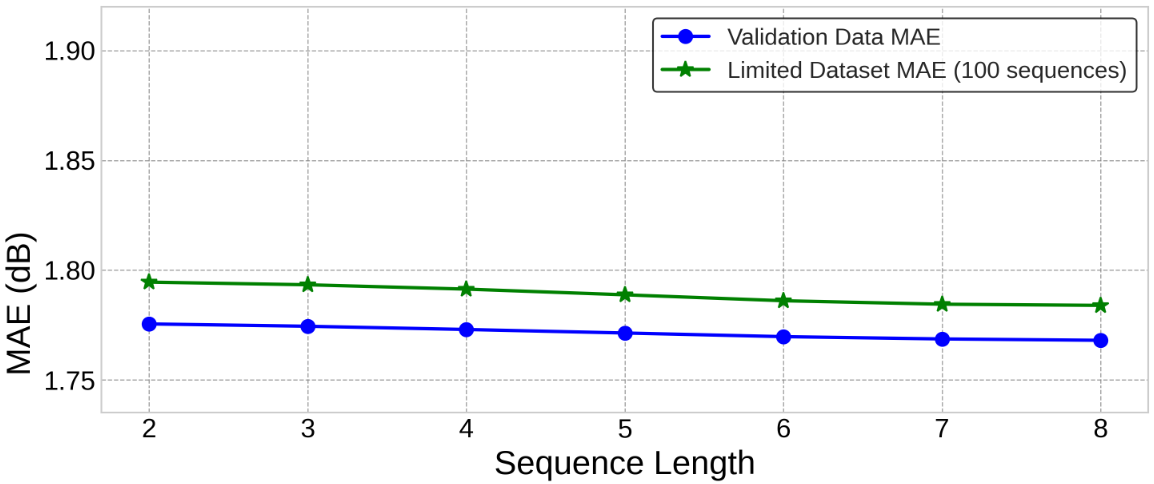}
        \vspace{-0mm} 
        \centerline{\footnotesize (a) MAE Comparison}
    \end{minipage}
    \vspace{2mm} 
    \begin{minipage}{1.0\columnwidth}
        \centering
        \includegraphics[width=0.95\linewidth]{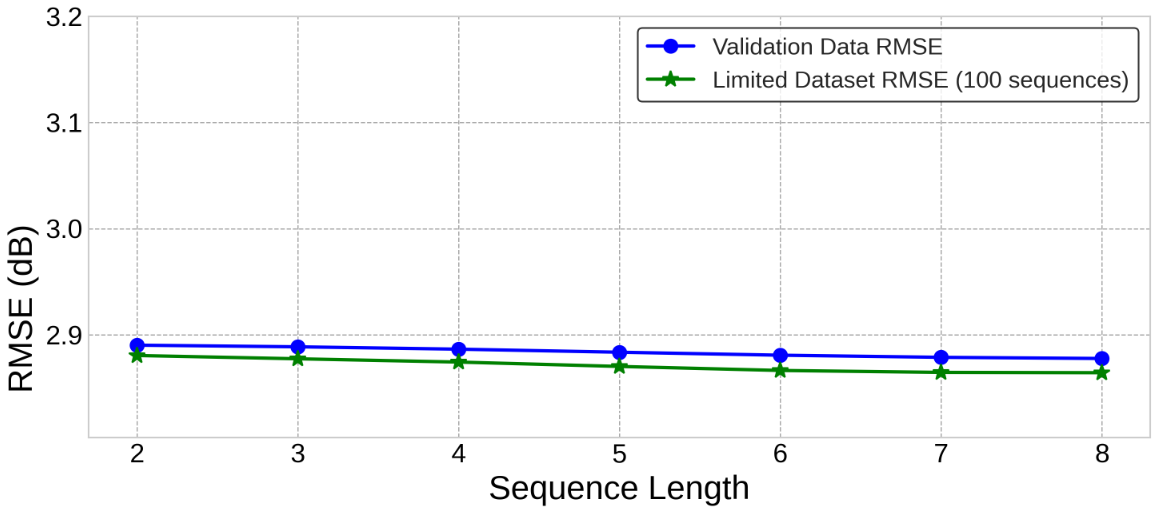}
        \vspace{-2mm}
        \centerline{\footnotesize (b) RMSE Comparison}
    \end{minipage}
    \vspace{-2mm} 
    \caption{Predictive performance of Evo-WISVA on the overall test dataset versus a limited 100-sample time-series dataset.}
    \label{fig:global_metrics}
    \vspace{-4mm} 
\end{figure}

Figure~\ref{fig:global_metrics} illustrate the model's Mean Absolute Error (MAE) and Root Mean Square Error (RMSE), for both the full validation dataset and the limited time-series subset. The results demonstrate that the model maintains a consistent level of performance, with only minor fluctuations, confirming its robust generalization capability.



The model's resilience to data volume variations is crucial for a proactive network management, as it must maintain predictive accuracy when operating on-the-fly with continuous data streams from a live network environment. The results confirm that the model's performance is not contingent on a large, diverse data sample at runtime, which is a key requirement for practical deployment.

\vspace{-4mm}
\subsection{Real-time Prediction Performance and Scalability Analysis}
A key advantage of the Evo-WISVA model is its computational efficiency, which is essential for real-time applications. The overall prediction time for a given task is the sum of the data transfer time and the model's forward pass time. The average per-frame prediction time is then the total time divided by the total number of frames processed.

\begin{equation}
T_{\text{overall}} = T_{\text{transfer}} + T_{\text{forward pass}}
\label{eq:total_time_eval}
\end{equation}

\begin{equation}
T_{\text{avg per frame}} = \frac{T_{\text{overall}}}{N_{\text{frames}}}
\label{eq:avg_frame_time_eval}
\end{equation}

\begin{figure}[!h]
    \centering
    \includegraphics[width=1.0\columnwidth]{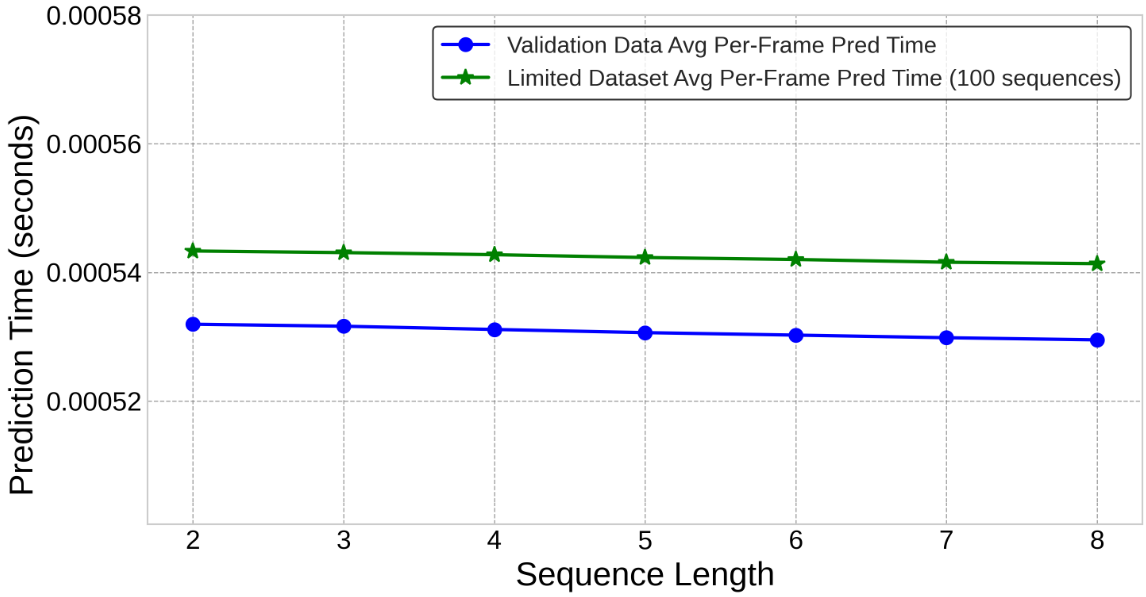}
    \caption{Average per-frame prediction time versus sequence length.}
    \label{fig:avg_per_frame_pred_time}
\end{figure}

Figure~\ref{fig:avg_per_frame_pred_time} visually demonstrates the relationship between sequence length and per-frame prediction time. The plot confirms our hypothesis of amortized computational cost, where the average per-frame time decreases as the sequence length increases. This indicates that the fixed overheads (like data transfer) are distributed over a larger number of frames, making the model more efficient for longer, continuous data streams.

\vspace{-6mm}
\subsection{Generalization to Increased Environmental Complexity}
\vspace{-2mm}
One of the most significant challenges in developing a wireless prediction system for a smart warehouse is handling the unpredictability introduced by dynamic environmental elements. To assess the robustness and generalization ability of the proposed Evo-WISVA model, we designed a transfer evaluation setting. The model was trained exclusively on a dataset containing a single moving shelf and subsequently tested on unseen datasets where 2, 4, 6, and 10 shelves were moving simultaneously. This evaluation is particularly important, as the number of moving shelves directly correlates with the degree of multipath interference and temporal variability in the wireless channel.

\begin{figure}[!h]
    \centering
    \includegraphics[width=1.0\columnwidth]{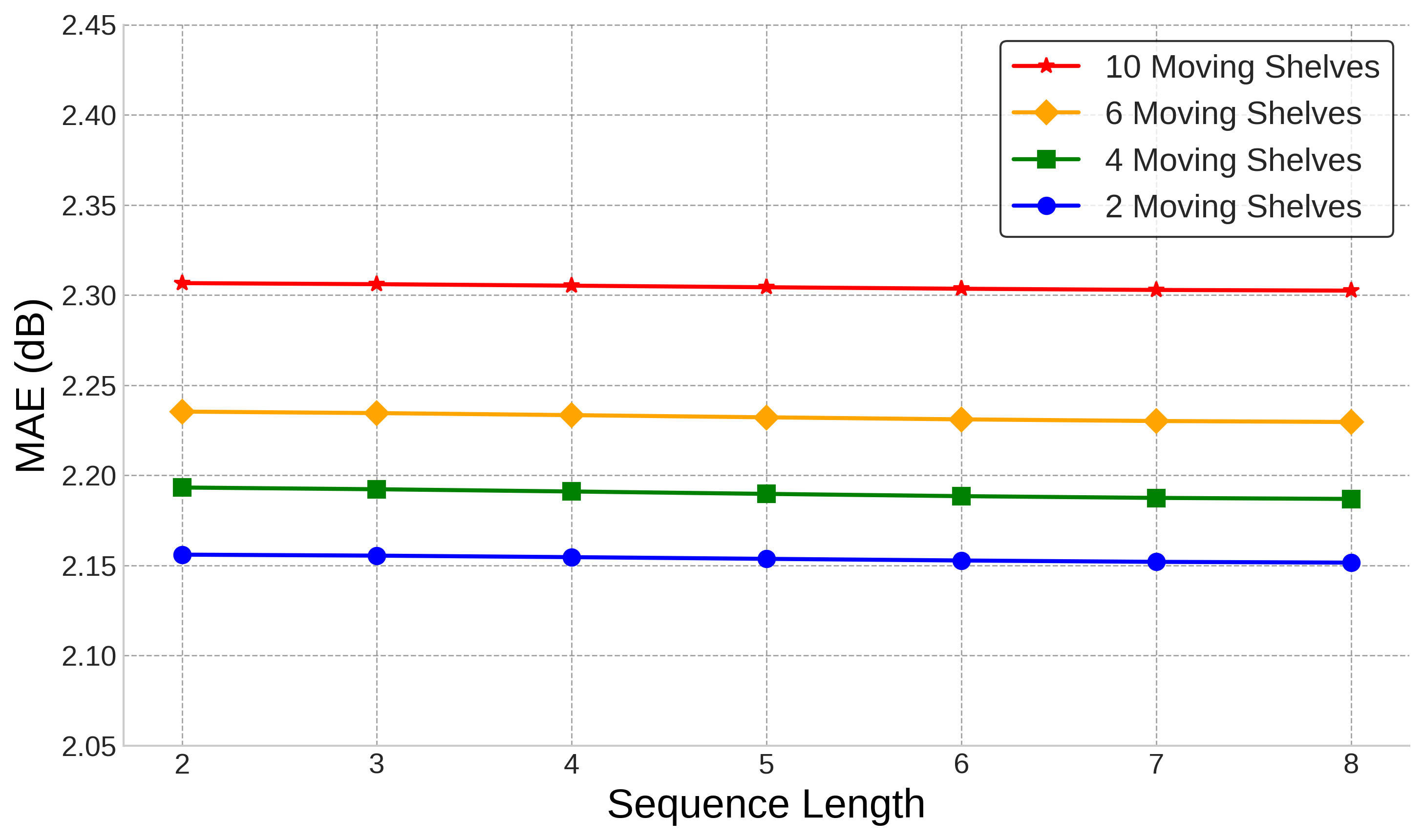}
    \caption{Evo-WISVA performance when tested on a dataset with increased dynamic complexity.}
    \label{fig:large_dynamic_mae}
\end{figure}

Figure~\ref{fig:large_dynamic_mae} presents the MAE across different sequence lengths for this scenario. A clear performance trend can be observed: as the number of moving shelves increases, the prediction error also gradually increases. For instance, when only two shelves were moving, the MAE remained close to $1.98$--$2.05$ dB across sequence lengths. In contrast, with ten shelves in motion, the MAE increased to approximately $2.20$--$2.52$ dB. Importantly, the error growth was incremental rather than catastrophic, suggesting that the model scales gracefully with environmental complexity.

These results highlight two critical insights. First, Evo-WISVA demonstrates strong generalization capacity despite being trained on a simplified scenario. The model successfully extrapolates to environments with significantly more dynamic interactions, preserving predictive fidelity even when the number of moving objects increased by an order of magnitude. Second, the relatively small increase in error values (on the order of $0.2$--$0.4$ dB in MAE when moving from two to ten shelves) indicates that the architecture is not merely overfitting to specific mobility patterns but instead captures the essential spatio-temporal structure of the wireless channel.

The memory-augmented VAE component proved essential for disentangling the stable physical properties of the warehouse layout from the stochastic, time-varying perturbations induced by moving shelves. Coupled with the ConvLSTM’s ability to model sequential dependencies, the architecture was able to preserve high accuracy under conditions that were far more complex than those encountered during training. This level of robustness is crucial for the deployment of predictive systems in real-world Industry~4.0 environments, where environmental dynamics cannot be exhaustively pre-encoded in training datasets.

\vspace{-4mm}
\section{Conclusion}
\vspace{-2mm}
In this paper, we introduced Evo-WISVA, a novel synergistic deep learning architecture for a Memory-Augmented Wireless Predictive System, specifically designed for dynamic SINR forecasting in Industry 4.0 warehouse environments. By integrating a memory-augmented Variational Autoencoder (VAE) with a ConvLSTM in an end-to-end jointly trained framework, our solution effectively addresses the challenges posed by mobile obstacles and complex RF propagation. We demonstrated that Evo-WISVA can generate accurate spatio-temporal SINR predictions while achieving significant computational acceleration, establishing it as a foundational technology for proactive wireless resource management and advanced predictive systems.

Industrial environments are inherently dynamic, rendering static network planning insufficient. The proposed model enables real-time foresight into wireless performance, allowing network operators to transition from reactive troubleshooting to proactive resource management. This represents a paradigm shift in industrial connectivity, as Evo-WISVA supports decision-making for dynamic beamforming, intelligent routing, and other adaptive network operations. By emulating complex wireless effects quickly and accurately as a neural surrogate, it reduces reliance on time-consuming physical simulations or measurements, further enhancing operational efficiency and reliability.

Future research directions include extending the model to multi-AP scenarios with dynamic handovers and interference management, integrating real-time sensor data (e.g., LiDAR-based obstacle detection) to improve prediction fidelity, and deploying the architecture on resource-constrained edge devices such as industrial gateways or autonomous mobile robots. Additionally, exploring the transferability of Evo-WISVA to other dynamic environments beyond warehouses will be key to establishing a generalizable framework for predictive industrial informatics.


\vspace{-4mm}
\bibliographystyle{IEEEtran} 
\bibliography{main}

\end{document}